\documentstyle[12pt]{article}
\begin{document}

\hfill PUPT-1729

\hfill hep-th/9710031

\vspace{1.0in}

\begin{center}

{\large\bf Notes on }

{\large\bf Heterotic Compactifications}

\vspace{1.0in}

Eric Sharpe \\
Physics Department \\
Princeton University \\
Princeton, NJ  08544 \\
{\tt ersharpe@puhep1.princeton.edu} \\

\vspace{0.5in}

\end{center}

In this technical note we describe a pair of results on heterotic
compactifications.  First, we give an example demonstrating
that the usual statement of the anomaly-freedom constraint
for perturbative heterotic compactifications (meaning,
matching second Chern characters) is incorrect for
compactifications involving torsion-free sheaves.
Secondly, we correct errors in the literature
regarding the counting of massless particles in heterotic
compactifications.

\begin{flushleft}
October 1997   
\end{flushleft}

\newpage

\section{Introduction}

Historically compactifications of heterotic string theory have been
extremely difficult to study.  Not only must one specify a Calabi-Yau
variety to compactify on, but one must also specify at least one sheaf.
Although in the last ten years we have come to a good understanding
of a large number of Calabi-Yau's, the physics community still does not
have a good grasp on compactifications that also involve sheaves.

There has been some limited progress in this direction, but it 
leaves much to be desired.  For example, methods to study bundles
on elliptic Calabi-Yau's \cite{wmf,wmf2,bbundle,donagi} have recently
been developed, but unfortunately this does not give any insight
into more general sheaves, or sheaves on Calabi-Yau's that are not
elliptic.  An older approach to the problem \cite{(02)} constructs
heterotic conformal field theories as infrared limits of gauged linear
sigma models \cite{phases}, but gives little understanding
of moduli space questions.

This paper is one of a set \cite{meallen} devoted to beginning to
put our understanding of heterotic compactifications on footing as
solid as that of type II compactifications.  In particular,
in this paper we correct two misconceptions in the
literature.  First we observe that the usual statement of anomaly
freedom in perturbative heterotic compactifications is incorrect.
Second, we correct misunderstandings in the literature regarding
the counting of massless particles in geometric compactifications
of heterotic string theory.

While our thesis advisor was reviewing
a final draft of this paper, the work \cite{ralph2} appeared,
which has overlap with the results in section~\ref{massless}.

\section{A Rapid Review of Heterotic Compactifications}

For a consistent perturbative compactification of either the
$E_{8}\times E_{8}$ or $Spin(32)/{\bf Z}_{2}$ heterotic string, in
addition to
specifying a Calabi-Yau $Z$ one must also specify
a set of Mumford-Takemoto semistable \cite{okonek,huybrechtslehn},
torsion-free sheaves\footnote{In this paper, by a locally free sheaf,
we mean a bundle.  A torsion-free sheaf is locally free up to
complex codimension two, and a reflexive sheaf is locally free
up to complex codimension three, on a smooth variety.  For more
information see  \cite{okonek,huybrechtslehn,meallen}.} 
$V_{i}$.
These sheaves must obey two constraints.  
One constraint\footnote{For example, for compactifications to four
dimensions, N=1 supersymmetry, on a Calabi-Yau $X$ one gets a 
D term in the low-energy
effective action proportional to $\langle X | \omega^{2} \cup c_{1}(V)
\rangle$.} can be written as
\begin{displaymath}
\omega^{n-1} \cup c_{1}( V_{i} ) \: = \: 0
\end{displaymath}
where $n$ is the complex dimension of the Calabi-Yau, and $\omega$ is
the K\"{a}hler form.  The other is an anomaly-cancellation condition
which,
if a single $V_{i}$ is embedded in each $E_{8}$, has historically been 
written as
\begin{displaymath}
\sum_{i} \, \left( c_{2}(V_{i}) \, - \, \frac{1}{2} c_{1}(V_{i})^{2} \,
\right) \: = \: c_{2}(TZ)
\end{displaymath}
In section~\ref{anomfree} we show that it is possible to have
a consistent perturbative heterotic compactification in which this constraint
is violated.

It was noted \cite{dmw}
that the anomaly-cancellation conditions can be
modified slightly by the presence of five-branes in the heterotic
compactification.  Let $[W]$ denote the cohomology class of the
five-branes, then the second constraint above is modified to
\begin{displaymath}
\sum_{i} \, \left( c_{2}(V_{i}) \, - \, \frac{1}{2} c_{1}(V_{i})^{2} \,
\right) \: + \: [W] \: = \: c_{2}(TZ)
\end{displaymath}
In this technical note we will only be concerned with 
perturbative heterotic compactifications.

Historically, for a long time the only perturbative heterotic
compactifications studied were those in which one took $V \, = \, TZ$,
the ``standard embedding."
This was done partly because more general compactifications are
more difficult to work with, and partly because it was believed more
general compactifications were destabilized by worldsheet instantons
\cite{xen}.  For the (0,2) models of \cite{(02)},
both difficulties have been overcome
\cite{(02),eva(02)}.

\section{Anomaly-free compactifications} \label{anomfree}

In this section we will give an example of an anomaly-free
perturbative heterotic compactification on a Calabi-Yau $Z$ 
involving a sheaf ${\cal E}$
of $c_{1}({\cal E}) = 0$ such that $c_{2}({\cal E}) \neq c_{2}(TZ)$.

Before we begin, we should explain how this is possible.
There are two ways one can check that a perturbative heterotic
compactification is anomaly-free.  One way is to check that the
worldsheet conformal field theory has no chiral anomalies.
The other way is through a constraint in the low-energy
supergravity of the form
\begin{equation} \label{dh}
d H \: = \: tr R \wedge R \: - \: tr F \wedge F
\end{equation}
Ordinarily both methods yield the same result: 
\begin{displaymath}
c_{2}({\cal E}) \: = \: c_{2}(TZ)
\end{displaymath}
However, there is a subtlety.  Constraint~(\ref{dh}) is only
sensibly defined when ${\cal E}$ is a bundle.  Only in that
case can one define a connection, and thereby make sense
of $F$.  When ${\cal E}$ is a more general coherent sheaf,
one can not define a connection, and so there is no way to
assign a meaning to $F$.  In that case, the only way to
check for anomalies is to study the worldsheet conformal field theory.
In particular, below we will give an example of an anomaly-free 
nonsingular conformal field
theory describing a torsion-free sheaf ${\cal E}$ in which 
$c_{1}({\cal E}) = 0$ and 
$c_{2}({\cal E})
\neq c_{2}(TZ)$.

The specific example
we will consider is of a sheaf ${\cal E}$ on
an elliptic threefold $Z$ (fibered over the Hirzebruch surface ${\bf F}_{1}$).
The conformal field theory is constructed using a gauged
linear sigma model \cite{phases}, in the form of a (0,2)
model of Distler, Kachru \cite{(02)}.  We shall assume
the reader is familiar with the material in \cite{phases,(02)}.

More precisely, the Calabi-Yau $Z$ is realized as a hypersurface
in an ambient space $X$ which is a ${\bf P}^{2}$ fibered over
${\bf F}_{1}$.  This ambient space $X$ can be described as the
quotient\footnote{The technically astute reader will recognize
that this space is a toric variety, and can be constructed as
a fan with edges
\begin{displaymath}
\nu_{u} = (1,0,0,0), \: \nu_{v} = (-1,-1,-6,-9), \: \nu_{w} = (0,1,0,0),
\: \nu_{s} = (0,-1,-4,-6)
\end{displaymath}
\begin{displaymath}
\nu_{x} = (0,0,1,0), \: \nu_{y} = (0,0,0,1), \: \nu_{z} = (0,0,-1,-1)
\end{displaymath}
}
\begin{displaymath}
X \: = \: \frac{ {\bf C}^{7} \, - \, S_{exc}}{
({\bf C}^{\times})^{3} }
\end{displaymath}
in other words, as homogeneous coordinates modded out by
${\bf C}^{\times}$ actions, where the homogeneous coordinates
$u$, $v$, $w$, $s$, $x$, $y$, $z$ have weights under the
${\bf C}^{\times}$ actions $\lambda$, $\mu$, $\nu$ as
\begin{center}
\begin{tabular}{c|cccc|ccc}
\, & $u$ & $v$ & $w$ & $s$ & $x$ & $y$ & $z$ \\ \hline
$\lambda$ & 1 & 1 & 1 & 0 & 6 & 9 & 0 \\
$\mu$ & 0 & 0 & 1 & 1 & 4 & 6 & 0 \\
$\nu$ & 0 & 0 & 0 & 0 & 1 & 1 & 1 
\end{tabular}
\end{center}
The ``exceptional set'' $S_{exc}$ is a set of points to be
omitted from ${\bf C}^{7}$ before quotienting.  This set
depends upon the precise phase of $X$, 
in the language of \cite{phases}, or more abstractly,
$S_{exc}$ determines the precise representative of the
birational equivalence class.  We will only consider the 
case that the Calabi-Yau hypersurface is a $K3$-fibration,
in which event $S_{exc}$ is
\begin{displaymath}
S_{exc} \: = \: \{ u = v = 0 \} \cup \{ w = s = 0 \} \cup
\{ x = y = z = 0 \}
\end{displaymath}

We will (naively) specify the sheaf ${\cal E}$ as the
(restriction to the hypersurface of a)
kernel of a short exact sequence, defined on 
$X$ as
\begin{equation} \label{ourex}
0 \: \longrightarrow \: {\cal E} \: \longrightarrow \:
{\cal O}(1,0,1)^{3} \oplus {\cal O}(1,1,0) \oplus
{\cal O}(24,3,14) \: \stackrel{ F_{a} }{ \longrightarrow }
\: {\cal O}(28,4,17) \: \longrightarrow \: 0
\end{equation}
In the present notation, $c_{1}( {\cal O}(a,b,c) ) = a D_{u} + 
b D_{z} + c D_{s}$, where for example\footnote{We are
maliciously failing to distinguish between divisor classes
and elements of $H^{2}({\bf Z})$.}
$D_{u} = \{ u = 0 \}$. 

We need to check that the (0,2) model defined above is anomaly-free
and defines an nonsingular conformal field theory.  To see that
it is anomaly-free, we compute the chiral anomalies associated
with each $U(1)$ in the linear sigma model. 

The linear sigma model contains a (0,2) chiral superfield for
each homogeneous coordinate, and in addition an auxiliary
superfield we shall label $p$, with charges under the $U(1)$'s
(which we shall label as $\lambda$, $\mu$, $\nu$, for obvious
reasons) as follows
\begin{center}
\begin{tabular}{c|cccc|ccc|c}
\, & $u$ & $v$ & $w$ & $s$ & $x$ & $y$ & $z$ & $p$ \\ \hline
$\lambda$ & 1 & 1 & 1 & 0 & 6 & 9 & 0 & -28 \\
$\mu$ & 0 & 0 & 1 & 1 & 4 &  6  & 0 & -17 \\
$\nu$ & 0 & 0 & 0 & 0  & 1 & 1 & 1 & -4 
\end{tabular}
\end{center}
In addition, the linear sigma model contains Fermi superfields
$\Lambda_{1}$, $\Lambda_{2}$, $\cdots$, $\Lambda_{5}$, $\Sigma$,
with $U(1)$ charges
\begin{center}
\begin{tabular}{c|ccccc|c}
\, & $\Lambda_{1}$ & $\Lambda_{2}$ & $\Lambda_{3}$ & $\Lambda_{4}$
& $\Lambda_{5}$ & $\Sigma$ \\ \hline
$\lambda$ & 1 & 1 & 1 & 1 & 24 & -18 \\
$\mu$ & 1 & 1 & 1 & 0 & 14 & -12 \\
$\nu$ & 0 & 0 & 0 & 1 & 3 & -3 
\end{tabular}
\end{center}
The (0,2) chiral superfields each contain a right-moving fermion,
and the Fermi superfields each contain a left-moving fermion.
Schematically, if we let $q_{+ i}^{(k)}$ denote the charge of the
$i$th right-moving fermion with respect to the $k$th $U(1)$,
and similarly for $q_{- i}^{(k)}$, then the anomaly cancellation
condition can be written as
\begin{displaymath}
\sum_{i} q_{+ i}^{(j)} q_{+ i}^{(k)} \: = \:
\sum_{i} q_{- i}^{(j)} q_{- i}^{(k)}
\end{displaymath}
for all pairs $j,k$.  It is straightforward to check that the
linear sigma model defined above is anomaly-free.
Ordinarily this result would imply\footnote{In fact, demanding
anomaly-freedom is somewhat stronger \cite{dgm}, as it (ordinarily) demands
matching Chern characters in each possible phase of the linear
sigma model.  In order to match Chern characters in each phase,
linear relations that exist between divisors can not be taken into
account.} 
$c_{2}({\cal E}) = c_{2}(TZ)$,
however we shall see momentarily that in the present case this is false.

In fact we have been quite naive in equation~(\ref{ourex}).
It is easy to check that each of the maps $F_{a}$ necessarily vanishes
at $x = z = 0$, so the short exact sequence shown is actually
not exact.  This problem and its solution were first described
in \cite{dgm}.  The sequence shown can be corrected to an
exact sequence by adding a term as
\begin{displaymath} 
0 \: \rightarrow \: {\cal E} \: \rightarrow \:
{\cal O}(1,0,1)^{3} \oplus {\cal O}(1,1,0) \oplus
{\cal O}(24,3,14) \: \rightarrow 
\: {\cal O}(28,4,17) \: \rightarrow \: 
{\cal O}_{S}(28,4,17) \: \rightarrow \: 0
\end{displaymath}
where ${\cal O}_{S}(28,4,17)$ is a skyscraper sheaf
localized over the subvariety $S = \{ x = z = 0 \}$.
Since $S$ is codimension two, ${\cal E}$ is a torsion-free
sheaf on $Z$.

Often when all of the maps $F_{a}$ in a short exact sequence
such as (\ref{ourex}) vanish, the associated conformal field
theory becomes singular.  In the linear sigma model,
this can be seen as a boson gaining a noncompact flat direction.
Here, however, this does not happen -- on the subvariety $S$,
none of the linear sigma model's bosons gain noncompact flat directions.
(This phenomenon was also first discussed in \cite{dgm}.)
This can be seen by examining the $D$ terms for the linear sigma model
describing this (0,2) model.  
The three $D$-term constraints\footnote{We are maliciously
failing to distinguish chiral superfields, their bosonic
components, and the vacuum expectation values of the bosonic
components.} are
\begin{eqnarray*} 
| u |^2 + | v |^2 + | w |^2 + 6 | x |^2 + 9 | y |^2 - 28 | p |^2
& = & r_{1} \\
| w |^2 + | s |^2 + 4 | x |^2 + 6 | y |^2 - 17 | p |^2 & = & r_2 \\
| x |^2 + | y |^2 + | z |^2 - 4 | p |^2 & = & r_3
\end{eqnarray*}
where the $r_{i}$ are Fayet-Iliopoulos terms.
Over $S = \{ x = z = 0 \}$ these constraint equations can be rewritten as
\begin{eqnarray} 
| u |^2  + |  v |^2 - | s |^2 + | p |^2 & = & r_1 - r_2 - 3 r_3 \\
| w |^2 + | s |^2 + 7 | p |^2 & = & r_2 - 6 r_3  \label{Dtermremix} \\
| y |^2 - 4 | p |^2 & = & r_3 
\end{eqnarray}
The $K3$-fibration phase is the region 
\begin{eqnarray*}
r_1 - r_2 - 3 r_3 & > & 0 \\
r_2 - 6 r_3 & > & 0 \\
r_3 & > & 0
\end{eqnarray*}
The only boson whose vacuum expectation value is in danger of
becoming unbounded over $S$ is $p$, and
it should be clear from equation~(\ref{Dtermremix})
that the vacuum expectation value of $p$ is bounded.

So far we have found an example of a conformal field theory
that is nonsingular and anomaly-free, describing a torsion-free
sheaf over a Calabi-Yau.  Although the conformal field
theory is anomaly-free, we now claim that $c_{1}({\cal E}) = 0$
and $c_{2}({\cal E}) \neq c_{2}(TZ)$.

The total Chern class of ${\cal E}$ can be calculated as
\begin{displaymath}
c( {\cal E} ) \: = \: \frac{ c \left[ {\cal O}(1,0,1)^{3} \oplus
{\cal O}(1,1,0) \oplus {\cal O}(24,3,14) \right] }{ c \left[
{\cal O}(28,4,17) \right]  }  c \left[ {\cal O}_{S}(28,4,17) \right]
\end{displaymath}
The skyscraper sheaf contributes to the second Chern class as
\begin{displaymath}
c_{2}({\cal E}) \: = \: c_{2}(TZ) \: - 6 D_{u} \cdot D_{z}
\: - \: 4 D_{s} \cdot D_{z} \: - \: D_{z}^{2}
\end{displaymath}
In particular, in the $K3$-fibration phase $H^{4}(X, {\bf Z})$ is generated by
$D_{u} \cdot D_{s}$, $D_{u} \cdot D_{z}$, $D_{s} \cdot D_{z}$, and 
$D_{z}^{2}$,
so this contribution is necessarily nonzero.

Finally, note that the subvariety $S = \{ x = z = 0 \}$ is located
on the Calabi-Yau hypersurface (in fact, is the section of the
elliptic fibration), and that for generic complex structure the
Calabi-Yau is smooth.  Thus, the restriction of ${\cal E}$ 
to the Calabi-Yau defines a torsion-free sheaf on the Calabi-Yau,
such that $c_{1}({\cal E}) = 0$ and $c_{2}({\cal E}) \neq c_{2}(TZ)$,
and the associated conformal field theory is anomaly-free and
nonsingular.

In principle, we should also check stability of ${\cal E}$,
but unfortunately except for a few consistency checks (see
\cite{ralph} for a recent discussion),
it is not known how to do this for (0,2) models, so we shall
simply ignore this difficulty.  Also, (0,2) models sometimes
suffer from certain additional poorly-understood anomalies
\cite{sdanom}.  In the event our particular example
suffered from either difficulty, we are completely confident other,
better behaved, 
examples 
could be found.

\section{Massless modes in heterotic compactifications} \label{massless}

As is well-known, (geometric) perturbative string compactifications have
massless modes corresponding to complex and Kahler deformations of the 
Calabi-Yau.  In (geometric) heterotic compactifications
there are additional charged and neutral massless modes,
which have been historically
mapped to certain sheaf cohomology groups \cite{old(02)}.  In this section
we will demonstrate that for heterotic compactifications
involving coherent sheaves which are not locally free, the
standard lore must be modified -- the sheaf cohomology groups
must be replaced with $\mbox{Ext}$ groups.  

We will assume the reader has
some basic familiarity with sheaf theoretic homological algebra,
as for example can be gained by reading the second appendix of
\cite{me1}.  To review briefly, for any coherent sheaves
${\cal E}$, ${\cal F}$, there are two sheaf-theoretic versions
of $\mbox{Ext}$.  One is a sheaf, known as $\mbox{(local) }
{\em Ext}^{i}({\cal E}, {\cal F})$.  The other is a group,
known as $\mbox{(global) } \mbox{Ext}^{i}({\cal E}, {\cal F})$.
They are related to one another by spectral sequences:  the group
$\mbox{(global) } \mbox{Ext}$ is the limit of either of the spectral sequences
with 
second level terms
\begin{eqnarray*}
E_{2}^{p,q}  & = & H^{p}( {\em Ext}^{q}({\cal E}, {\cal F})) \\
E_{2}^{' p,q} & = & H^{q}( {\em Ext}^{p}({\cal E}, {\cal F}))
\end{eqnarray*}

First, it is often stated in the physics literature that deformations
of a sheaf ${\cal E}$ are in one-to-one correspondence with elements
of the sheaf cohomology group $H^{1}( {\em End } {\cal E})$.
However, this statement is strictly correct only when
${\cal E}$ is locally free.  More generally, when ${\cal E}$ 
is some arbitrary coherent sheaf, deformations of ${\cal E}$
are in one-to-one correspondence with elements of
(global) $\mbox{Ext}^{1}({\cal E},{\cal E})$.

The group $\mbox{Ext}^{1}({\cal E},{\cal E})$ receives
contributions from both $H^{1}({\em End } {\cal E})$
as well as $H^{0}( {\em Ext}^{1}( {\cal E}, {\cal E}))$.
When ${\cal E}$ is locally free, the sheaf ${\em Ext}^{1}({\cal E},-)
 = 0$, and so in this special case $\mbox{Ext}^{1}({\cal E},{\cal E})
= H^{1}({\em End } {\cal E})$.

This fact has consequences for the other massless modes.
For example, consider embedding a rank 3 torsion-free sheaf
${\cal E}$ in an $E_{8}$, breaking it to $E_{6}$.  All massless modes
in the compactification arise from deformations of the
ten-dimensional adjoint $E_{8}$ vector, so as we have
the group-theoretic decomposition of the of the {\bf 248} of
$E_{8}$ into representations of $SU(3) \times E_{6}$ as
\begin{displaymath}
{\bf 248} \: = \: ({\bf 1}, {\bf 78}) \oplus ({\bf 3}, {\bf 27})
\oplus (\overline{{\bf 3}}, \overline{{\bf 27}}) \oplus
({\bf 8}, {\bf 1})
\end{displaymath}
we have massless scalars in $E_{6}$ representations as 
\begin{center}
\begin{tabular}{c|c}
Group & $E_{6}$ representation \\ \hline
$\mbox{Ext}^{1}({\cal O}, {\cal O})$ & ${\bf 78}$ \\
$\mbox{Ext}^{1}({\cal O}, {\cal E})$ & ${\bf 27}$ \\
$\mbox{Ext}^{1}({\cal E}, {\cal O})$ & $\overline{{\bf 27}}$ \\
$\mbox{Ext}^{1}({\cal E}, {\cal E})$ & ${\bf 1}$ \\
\end{tabular}
\end{center}
(Note that $\mbox{Ext}^{1}({\cal O},{\cal O}) = h^{1,0} = 0$
on a Calabi-Yau $n$-fold for $n > 1$, so in these cases there are
no scalars in the ${\bf 78}$ (only a vector).)
In the existing literature it is often incorrectly claimed these massless
modes are counted by sheaf cohomology groups; here we see the
correct counting involves $\mbox{Ext}$ groups instead.
(In the special case that ${\cal E}$ is locally free,
the $\mbox{Ext}$ groups simplify to sheaf cohomology groups.)

Many of these $\mbox{Ext}$ groups are equal to sheaf
cohomology groups for arbitrary coherent ${\cal E}$.  
For example\footnote{This is because
${\em Ext}^{i}( {\cal O}_{X}, - )$ is the right derived functor
of ${\em Hom}( {\cal O}_{X}, - )$, which is the identity functor.
By contrast, ${\em Hom}( -, {\cal O}_{X})$ is not the identity
functor, and in general its right derived functor ${\em Ext}^{i}(-,
{\cal O}_{X})$ is nonzero for $i > 0$.} 
\cite[section III.6]{hartshorne}, 
on any variety $X$,
for all coherent sheaves ${\cal E}$, 
\begin{displaymath}
\mbox{(local) } {\em Ext}^{i}( {\cal O}_{X}, {\cal E}) \: = \:
\left\{ \begin{array}{ll}
        {\cal E} & i = 0 \\
        0 & i > 0
        \end{array} \right.
\end{displaymath}
so in particular 
\begin{displaymath}
\mbox{(global) } \mbox{Ext}^{i} ( {\cal O}_{X}, {\cal E} ) \: \cong
\: H^{i} ( X , {\cal E} ) \: \forall i \geq 0
\end{displaymath}

Another useful fact \cite[section III.6]{hartshorne} is that
if ${\cal L}$ is any locally free sheaf and ${\cal F}$, ${\cal G}$
are any coherent sheaves, then
\begin{displaymath}
\mbox{(local) } {\em Ext}^{i} ( {\cal F} \otimes {\cal L}, {\cal G})
\: \cong \: {\em Ext}^{i} ( {\cal F}, {\cal L}^{\vee} \otimes {\cal G} )
\: \cong \:  {\em Ext}^{i} ( {\cal F}, {\cal G} ) \otimes {\cal L}^{\vee}
\end{displaymath}
and so
\begin{displaymath}
\mbox{(global) } \mbox{Ext}^{i}({\cal F} \otimes {\cal L}, {\cal G})
\: \cong \: \mbox{Ext}^{i} ( {\cal F}, {\cal L}^{\vee} \otimes {\cal G})
\end{displaymath}

In \cite{old(02)} it was noted that particle-antiparticle
duality was realized by Serre duality, a fact which we
expect to continue to hold even for non-locally-free sheaves.
For locally free sheaves ${\cal E}$, Serre duality on an
$n$-dimensional projective Cohen-Macaulay\footnote{
Examples of Cohen-Macaulay varieties include toric varieties
\cite{fulton} and Calabi-Yau's in dimensions less than 4 \cite{dave}.}
variety is the statement
\begin{equation} \label{oldserre}
H^{i}({\cal E}) \: \cong \: H^{n-i}({\cal E}^{\vee} \otimes \omega)^{\vee}
\end{equation}
where $\omega$ is the dualizing sheaf (equal to the canonical
bundle when the variety is smooth).
For more general coherent sheaves the correct statement\footnote{
The technically astute reader will recognize this as a special
case of the Yoneda pairing
\begin{displaymath}
\mbox{(global) } \mbox{Ext}^{p}({\cal F}, {\cal G}) \otimes
\mbox{Ext}^{q}({\cal G},{\cal H}) \: \rightarrow \:
\mbox{Ext}^{p+q}({\cal F},{\cal H})
\end{displaymath}
valid for all coherent ${\cal F}$, ${\cal G}$, and ${\cal H}$.
In fact, for the special case of coherent sheaves ${\cal E}$,
${\cal F}$ on a smooth projective variety, Serre duality can
be generalized \cite{rf,mukai} to the statement
\begin{displaymath}
\mbox{(global) } \mbox{Ext}^{i}({\cal E}, {\cal F}) \:
\cong \: \mbox{Ext}^{n-i}({\cal F}, {\cal E} \otimes \omega)^{\vee}
\end{displaymath}
}
of Serre duality on a projective Cohen-Macaulay variety 
is \cite[section III.7]{hartshorne}
\begin{displaymath}
\mbox{Ext}^{i}(  {\cal O}_{X}, {\cal E}) \: \cong \:
\mbox{Ext}^{n-i}( {\cal E}, \omega)^{\vee}
\end{displaymath}
and in particular on a Calabi-Yau variety this is simply
\begin{equation} \label{trueserre}
\mbox{Ext}^{i}( {\cal O}_{X}, {\cal E}) \: \cong \:
\mbox{Ext}^{n-i}( {\cal E}, {\cal O}_{X} )^{\vee}
\end{equation}

We should point out that this correction solves a problem associated
with any attempt to hypothesize (0,2) mirror symmetry for torsion-free
sheaves.  
In principle, if $X$, $Y$ are two Calabi-Yau's with sheaves
${\cal E}$, ${\cal F}$, respectively, such that each pair defines
an anomaly-free heterotic conformal field theory, then if
($X$,${\cal E}$), ($Y$,${\cal F}$) are (0,2) mirror pairs
then  
\begin{displaymath}
\mbox{dim } \mbox{Ext}^{i}_{X}({\cal E},{\cal O}_{X}) \: = \:
\mbox{dim } \mbox{Ext}^{i}_{Y}({\cal O}_{Y},{\cal F})
\end{displaymath}
or, equivalently (by Serre duality),
\begin{displaymath}
\mbox{dim } \mbox{Ext}^{i}_{X}({\cal O}_{X},{\cal E}) \: = \:
\mbox{dim } \mbox{Ext}^{i}_{Y}({\cal F},{\cal O}_{Y})
\end{displaymath}
Now, if we did not know that sheaf cohomology groups should
be replaced with $\mbox{Ext}$ groups, and 
incorrectly applied the duality in equation~(\ref{oldserre}), 
then the analogous
two statements 
\begin{eqnarray*}
\mbox{dim } H^{i}(X, {\cal E}^{\vee}) & = & 
\mbox{dim } H^{i}(Y, {\cal F}) \\
\mbox{dim } H^{i}(X, {\cal E}) & = & 
\mbox{dim } H^{i}(Y, {\cal F}^{\vee})
\end{eqnarray*}
would only be equivalent  
if 
${\cal E}$ and ${\cal F}$ were both reflexive sheaves.
In general, we would have two inequivalent
possible formulations of (0,2) mirror symmetry.  By recognizing
that massless modes are counted with $\mbox{Ext}$ groups
rather than sheaf cohomology groups, this ambiguity is resolved.

In passing we should probably mention a subtlety associated
with the Dirac index, which counts the generation number in
heterotic compactifications.  Strictly speaking, the Dirac
operator is only defined when coupled to a locally free sheaf,
because only then can we define a connection and thereby
make sense out of $D = \partial + A$.
As the Dirac operator is only defined when coupled to a 
locally-free sheaf, strictly speaking we can only define the
Dirac index when coupled to a locally free sheaf. 
So, how can we compute the generation number when working
with more general torsion-free sheaves?
Instead of computing the Dirac index to find the
generation number, one can also count
particles directly.  In general by counting chiral fermions
one finds that the generation number is of the form
\begin{displaymath}
\sum_i \, (-)^i \, \mbox{dim } \mbox{Ext}^i_X( {\cal E},
{\cal F})
\end{displaymath}
for torsion-free sheaves ${\cal E}$ and ${\cal F}$.
The Riemann-Roch theorem for general coherent sheaves \cite{mukai2}
is
\begin{displaymath}
\sum_i \, (-)^i \, \mbox{dim } \mbox{Ext}^i_X( {\cal E}, {\cal F})
\: = \: \langle X \, | \, ch({\cal E})^* \cup ch({\cal F}) \cup
td(TX) \rangle
\end{displaymath}
where by $ch({\cal E})^*$ we mean that all cohomology elements of
degree $4n+2$ (for some $n$) in $ch({\cal E})$ should be multiplied
by $-1$.  (For example, if ${\cal E}$ were a bundle, then $ch({\cal E})^*
= ch({\cal E}^{\vee})$.)  In particular, we recover the result
for the generation number that one would have naively guessed.
For example, for a compactification on a threefold $X$ involving a torsion-free
sheaf ${\cal E}$ with $c_1({\cal E}) = 0$, the generation number
is proportional to $\langle X | c_3({\cal E}) \rangle$,
formally identical to the result for the special case ${\cal E}$ is a bundle.

\section{Conclusions}

In this paper we have corrected two common misconceptions
regarding perturbative compactifications of heterotic string
theory.  We have demonstrated that the usual statement of the anomaly
freedom constraint is incorrect in general, and we have also corrected
a common miscounting of massless particles in heterotic string theory.

Far more work remains to be done before our understanding of heterotic
compactifications begins to approach our understanding of type II
compactifications.  At this point the physics community does not
even have a good understanding of classical 
results, much less any understanding of quantum 
corrections.  Significant progress in understanding the classical
physics will be reported in \cite{meallen}.

\section{Acknowledgements}

We would like to thank R. Friedman, D. Morrison, and
E. Witten for useful discussions.

\end{document}